\documentclass[prl,aps,twocolumn,showpacs,twoside,10pt]{revtex4-1}
\usepackage{graphicx}
\usepackage{times}

\newcommand{\kr}{k_\mathrm{r}}
\newcommand{\Tt}{T_\mathrm{turn}}
\newcommand{\llat}{\lambda_\mathrm{Lat}}
\newcommand{\qm}{q_\mathrm{max}}
\newcommand{\Erec}{E_\mathrm{rec}}

\newcommand{\om}[1]{\nu_{#1}}
\newcommand{\omm}{\nu_\mathrm{mod}}
\newcommand{\nurm}[1]{\nu_\mathrm{#1}}
\newcommand{\omT}{\omega_\mathrm{trap}}
\newcommand{\unit}[2]{#1~\mathrm{#2}}

\newcommand{\Vlat}{V_\mathrm{Lat}}

\begin{document}
\title{Pump-probe coupling of matter wave packets to remote lattice states}

\author{Jacob F. Sherson}
\email{sherson@phys.au.dk}
\author{Sung Jong Park}
\altaffiliation{Korea Research Institute of Standards and Science, Yuseong, Daejeon
305-340, Korea}
\author{Poul Pedersen}
\author{Nils Winter}
\author{Miroslav Gajdacz}
\author{Sune Mai}
\author{Jan Arlt}
\affiliation{Danish National Research Foundation Center for Quantum Optics QUANTOP, Department of Physics and Astronomy, University of Aarhus, DK-8000 Aarhus C, Denmark}

\date{\today}
\begin{abstract}
The coherent manipulation of wave packets is an important tool in many areas of physics. We demonstrate the experimental realization of quasi-free wave packets of ultra-cold atoms bound by an external harmonic trap. The wave packets are produced by modulating the intensity of an optical lattice containing a Bose-Einstein condensate. The evolution of these wave packets is monitored \textit{in-situ} and their reflection on a band gap is observed. In direct analogy with pump-probe spectroscopy, a probe pulse allows for the resonant de-excitation of the wave packet into localized lattice states at a long, controllable distance of more than 100 lattice sites from the main component.
This coherent control mechanism for ultra-cold atoms thus enables controlled quantum state preparation, opening exciting perspectives for quantum metrology and simulation.
\end{abstract}
\pagestyle{plain}
\maketitle


Wave packets - non-stationary superpositions of energy states -  have held a special place in physics ever since Schr\"{o}dinger
showed that a superposition of states can form a spatially localized entity evolving similar to a classical particle~\cite{Schroedinger-1928}. Despite being ubiquitous in nature, wave packets remained an elusive concept for decades. The advent of ultra-short laser pulses has, however, enabled a multitude of observations of wave packet phenomena
in e.g. highly excited atomic Rydberg states~\cite{Zoller-1991,Maeda-2009}, molecular states~\cite{Garraway1995,Brabec2000,Salzmann2008,Krausz2009}, and semiconductor systems~\cite{Ulbricht2011}.

One of the main applications of wave-packets has been the investigation of complex quantum systems using so-called pump-probe spectroscopy~\cite{Garraway1995,Krausz2009,Ulbricht2011}. An initial short pump pulse is used to create a wave packet in an excited state. Subsequently, the wave packet is allowed to evolve for a variable duration. Finally, a probe pulse interrogates this dynamic evolution by transferring it to a chosen final state. This process can thus provide information about coherent dynamics in the excited state~\cite{Salzmann2008,Goulielmakis2010}, but it can equally well be used for e.g. interferometry~\cite{Scherer1991} or for the preparation of a desired final state~\cite{Herek1994}.

In parallel, the precise control of matter waves in Bose-Einstein condensates (BECs) has enabled a new field of wave packet manipulation~\cite{Cronin2009}. For instance, Bragg diffraction~\cite{Kozuma-1999} has been utilized to split and recombine wave packets interferometrically~\cite{Denschlag-2000} and
freely propagating atomic wave-packets have been coupled out of a trapped BEC~\cite{Mewes1997}.
These implementations, however, lack the structural complexity due to the combination of bound and continuum states present in e.g. molecular and semiconductor systems.

In this letter we present the experimental realization of quasi-free wave packets of ultra-cold atoms
created from an initial bound state.
We produce these wave packets from an ultra-cold cloud by briefly modulating the intensity of an optical lattice in the presence of an external harmonic trap. The ensuing oscillatory motion of the wave packets is observed {\it in-situ} and shows only limited spreading. In an important application of this new phenomenon we couple the wave packets into localized lattice states~\cite{Viverit-2004,Ott-2004} separated from the original cloud by more than 100 lattice sites using a pump-probe approach directly analogous to the method discussed above.


\begin{figure}[tbp]
\includegraphics[width=8cm]{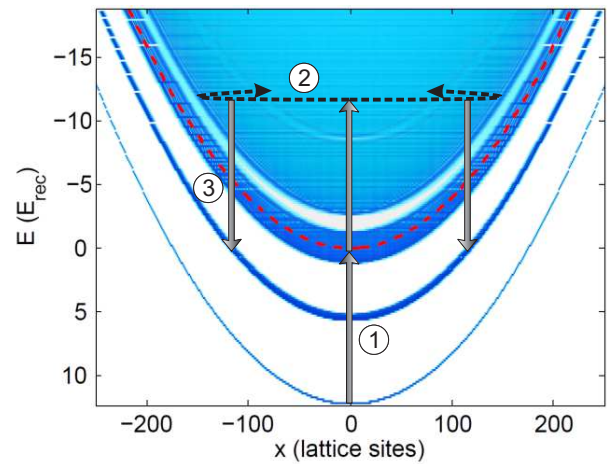}
\caption{(Color online) Spectrum of the 1D single particle Schr\"{o}dinger equation for a combined harmonic ($\omT = 2\pi\cdot37.9$~Hz) and periodic potential at a lattice depth $s = 16$. The dashed red line indicates the maximum of the periodic potential. Using a short pulse of lattice amplitude modulation, quasi-continuum wave packets are created~(1), which oscillate in the
potential~(2). An appropriately timed second modulation pulse can de-excite the wave packets into localized states~(3).}
\label{fig:schroedinger}
\end{figure}

The production of wave packets and the population of outlying localized states can best be visualized using  the population density of the energy eigenstates. Figure~\ref{fig:schroedinger} shows the calculated 1D energy spectrum of the single particle Schr\"odinger equation in the combined harmonic and optical lattice potential~\cite{Viverit-2004,Ott-2004}. It clearly shows pseudo-bands of allowed energies caused by the optical lattice, which are shifted according to the spatial dependence of the harmonic trap. Due to the site-to-site energy shift induced by the harmonic confinement, tunneling is inhibited outside the trap center in the bound bands, resulting in so-called localized states. Above the lattice threshold (red dashed line) a quasi-continuum of delocalized eigenstates of the combined potential is formed.

To realize this situation, the depth of the 1D optical lattice $\Vlat$ is chosen such that the second band is just barely trapped in the lattice, while higher bands are not. A harmonic modulation of the lattice depth with a frequency $\omm$ close to the transition frequency from the ground state to the second excited band, $\om{02}$, can induce transitions between these bands~\cite{Denschlag2002} and further  into the quasi-continuum states that exist above the threshold of the lattice.

In the following, we show experimentally that a short pulse of lattice modulation can induce this two-phonon transition into the quasi-continuum. This creates wave packets of narrow quasi-momentum distribution which propagate outwards along the direction of the optical lattice. We also show that detailed knowledge of the dynamics can be used to induce a resonant de-excitation into the outlying localized states analogous to pump-probe spectroscopy. These processes are schematically indicated in Fig.~\ref{fig:schroedinger}.


Our experimental system~\cite{Bertelsen-2007} consists of a BEC in a harmonic magnetic potential and a one-dimensional optical lattice. A BEC of $^{87}$Rb atoms in the $5^2 S_{1/2} |F=2, m_F=2\rangle$ state is prepared in a quadrupole-Ioffe configuration (QUIC) trap~\cite{Esslinger-1998}. After condensation, the magnetic potential is considerably relaxed to obtain trap frequencies of $12.3$~Hz and $37.9$~Hz in the axial and radial direction respectively.
Pure condensates of approximately $3 \times 10^5$ atoms, are then loaded adiabatically into the vibrational ground state of a 1D optical lattice.
The optical lattice is formed along the vertical axis (one of the radial directions of the trap) by a retro-reflected laser beam at a wavelength of $\llat = 914$~nm with a $1/e^2$ waist of 120 $\mu$m. For all  experiments the lattice depth was $s\equiv \Vlat/\Erec=16$, where $\Erec=h^2/(2m\llat^2)$.


\begin{figure}[tbp]
\includegraphics[width=8cm]{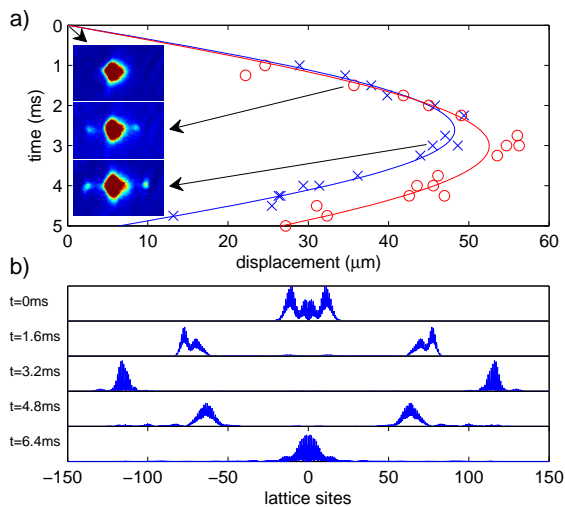}
\caption{(Color online) Propagation of the wave packet after excitation with a $500~\mu$s modulation pulse. a) Position of the upper ($\times$) and the lower ($\circ$) wave packet as a function of the evolution time after modulation. b) Wave packet dynamics obtained from a numerical simulation (see text).}
\label{fig:wavepacket}
\end{figure}

To produce the wave packets experimentally, a short ($\unit{500}{\mu s}$), large amplitude ($0.3\Vlat$) 
lattice modulation pulse is applied at a modulation frequency $\nurm{mod} = \unit{31}{kHz}$ close to $\nurm{02} = \unit{30.5}{kHz}$. Subsequently, images of the resulting clouds are taken \textit{in-situ} following a variable evolution time. Two wave packets, traveling in each lattice direction away from the original cloud  are observed as shown in Fig.~\ref{fig:wavepacket}~a)~(inset)~\footnote{All experimental images ware rotated 90$^{\circ}$ counter-clockwise for clarity.}.
The wave packets each  contain roughly 2 \% of the total atoms and do not show considerable spreading as they travel to the turning point, are reflected, and return. Due to gravity the dynamics of the wave packet traveling upwards is slightly modified compared to the one traveling downwards. For the quantitative analysis in the following we therefore use the upper component only. Figure~\ref{fig:wavepacket}~a) shows the distance of the wave packet from the center of the main component as a function of the evolution time during the first half oscillation period. It reveals a time to reach the turning point of roughly $\Tt=2.6$~ms ($\Tt=3.0$~ms) for the upper (lower) peaks. We stress that this observation is clearly inconsistent with harmonic oscillation at the trap frequency $\omT=2\pi\cdot37.9$~Hz from which we would expect a turning point at $6.6$~ms.

An intuitive understanding of the process can be obtained from an analysis based on Fig.~\ref{fig:schroedinger} and the conventional band structure represented in the first Brillouin zone.
The modulation pulse transfers a small fraction of the ground state population via a two-phonon excitation to a superposition of quasi continuum states,
determined by the individual coupling matrix elements.
In the band picture, this is equivalent to
a transfer into the fourth band for the chosen lattice depth.
Energy conservation dictates a transfer into states with opposite quasi-momentum $\pm\hbar \qm$, where $4\leq \qm/\kr\leq5$ and $\kr=2\pi/\llat$. The wave packets
exchange their initial kinetic energy for potential energy by moving outward in the harmonic potential while rolling down the appropriate bands.
As the first significant band gap is encountered, they inverse their direction due to a Bragg reflection.

This interpretation directly allows for a simple calculation of the expected behavior of the wave packets.
Treating the quasi-momentum $\hbar q$ as the momentum $p$ of a classical particle moving in a harmonic trap (i.e. $p(t)=\hbar\qm\cos(\omT t)$),
the times at which the quasi-momentum passes the avoided crossings between bands  $n$ and $(n-1)$  at $q=n k_r$  are
$t_n = \arccos\left( \frac{n \kr}{\qm}\right)/\omT$
for $0 \leq n<4$. We obtain $t_3 = \unit{3.0}{ms}$  for $\qm\sim4\kr$ in good agreement with the experimental result and therefore interpret the reflection of the wave packet as a Bragg-reflection on the gap between the second and the third band (6-photon Raman process).
 A calculation of the Landau Zener tunneling probability at the $n=2$ to $n=3$  band gap supports this, since it is essentially zero at $s=16$.

For a more quantitative understanding, we have also performed a single-particle numerical simulation of the process by solving the time dependent Schr\"{o}dinger equation using the rotating wave approximation.
We do this by splitting the spectrum into three parts: the zeroth band, the first and second excited band
manifold, and finally the continuum states. Coupling between the individual states of
separate manifolds is introduced according to their coupling matrix elements. Allowing for this dynamics to evolve for the
duration of the modulation pulse we obtain the excited wave packet as a  superposition of
eigenstates. The ensuing wave packet dynamics, dictated by the relative phase evolution of
these states, is shown in Fig.~\ref{fig:wavepacket}~b) and reveals a turning point at 3.2~ms.
Our simulation neglects interactions and since the mean field energy tends to flatten the effective potential we expect the inclusion of interactions
to decrease $\Tt$ and thus give better agreement with the experiment. Furthermore,
the simulation starts in the absolute single particle ground state, since interactions are neglected. We found that the simulation had to be performed at a relatively low modulation frequency (28 kHz) to obtain agreement with experiment and attribute this to the interaction induced broadening of the true initial state.


\begin{figure}[tbp]
\includegraphics[width=8cm]{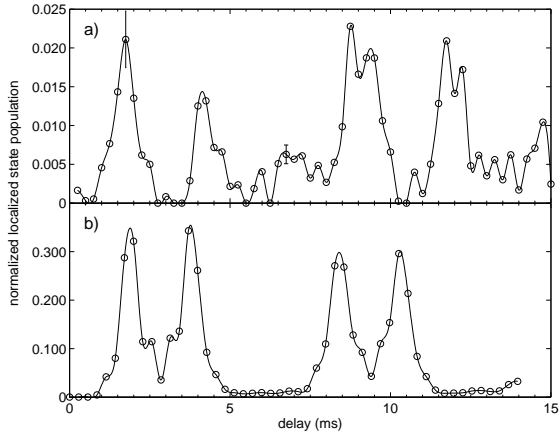}
\caption{Time dependence of the pump-probe spectroscopy. a) Population in the localized states as a function of the delay between the $500~\mu$s pump and probe pulses ($\omm=32$~kHz). The double peaked structure corresponds to the passage of the wave packet through the resonance region. Representative error bars are included at $t=1.75$ ms and at $t=6.75$ ms. b) Numerical simulation of the pump-probe experiment (see text).}
\label{fig:pumpProbeExp}
\end{figure}

The production process of wave packets described above bears strong resemblance to the pump pulse in a pump-probe experiment. We therefore illustrate the dynamical control of these wave packets by coupling them into outlying lattice states with a second, so-called probe, pulse. As mentioned above, these localized states appear since tunneling in the lattice is inhibited outside the trap center due to the large site-to-site energy shift. These states were predicted theoretically in Ref.~\cite{Viverit-2004} and experimentally observed in Ref.~\cite{Ott-2004}. Here, we show the first dynamic population and  \textit{in-situ} detection of such states. Unlike previous experiments, our method allows for the transfer of atoms into states that are separated by large distances ($\sim100$ sites) from the original cloud.
Furthermore, our transport mechanism does not simultaneously broaden the spatial distribution unlike tunneling based long distance transport~\cite{Alberti2009}. It  also allows for the coherent transfer of a controllable fraction of the cloud in contrast to conventional transport using accelerated optical lattices~\cite{Schmid2006}.

The condition for the population of these stationary states can be understood by returning to Fig.~\ref{fig:schroedinger}. For a given modulation frequency $\omm$, the oscillating wave packets after some evolution time have a spatial overlap with localized states at an energy difference of $h\omm$. At this time a brief modulation pulse will induce coupling to these states in correspondence with the Franck-Condon principle for transitions between vibrational states.

Experimentally, we investigate the transfer to localized states by adding a second $500~\mu$s amplitude modulation pulse after variable evolution times. The resulting population is shown in Fig.~\ref{fig:pumpProbeExp}~a).
We have verified that the states are indeed stationary, by checking that their population is independent of an additional hold time in the lattice before \textit{in-situ} imaging. 
We observe a double peaked resonance structure corresponding to the first passage and subsequent return of the wave packet following the reflection. In addition, we observe a second double peaked resonance after approx.~$10$~ms, corresponding to the wave packet that has completed $3/4$ of an oscillation after originally traveling downwards.
Figure~\ref{fig:pumpProbeExp} shows that roughly 2\% of the total population is observed in each localized peak, consistent with
a transfer of the majority of the wave packet population. A determination of the precise transfer
efficiency and the optimization of the process will be a topic of future investigation.

Figure~\ref{fig:pumpProbeExp}~b) shows the numerical simulation of the pump-probe dynamics obtained by adding a second modulation pulse to the  simulation discussed above.
We extract the bound state population and plot only the contribution from the outlying positions. The correspondence with the experimental data for early times is remarkable, which clearly confirms our detailed understanding of the process. Since the numerical simulation is performed at 28 kHz instead of 32 kHz the oscillation is slightly faster, which becomes evident as a clear shift of the second double peak.

\begin{figure}[t]
\centering
\includegraphics[width=8cm]{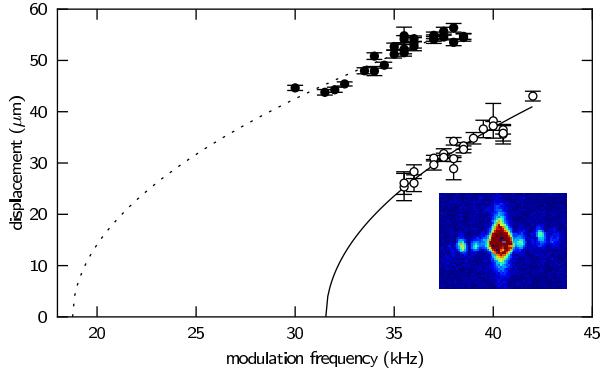}
\caption{(Color online) Positions of the populated localized states as a function of lattice modulation frequency. The fits correspond to the expected square root behavior (see Eq.~(\ref{eq:sqrtNum})).}
\label{fig:transp_outlier_pos_vs_freq}
\end{figure}

Finally, the position of the  localized states was systematically investigated by choosing a modulation duration of $10$~ms for various modulation frequencies below and above the transition frequency $\om{02}$ (see Fig.~\ref{fig:transp_outlier_pos_vs_freq}). 
We have verified that the duration is long enough to reach a steady state population.
The  position clearly moves outwards as the frequency is increased and as the resonance is crossed a new set of close lying components appear.
The inset of Fig.~\ref{fig:transp_outlier_pos_vs_freq} shows an image recorded at a modulation frequency of $36$~kHz, displaying the main cloud and an inner and outer set of localized states.

This behavior can, again, be understood based on the spectrum shown in Fig.~\ref{fig:schroedinger}. At  modulation frequencies $\omm\geq\om{02}$ population can be coupled both to the first and the second band, thus resulting in the two observed sets of populated outlying states. The position of each of these states, $x_i$, can be obtained by considering the energy balance between the modulation, the band, and the harmonic potential energies
\begin{equation}
2h\omm=  h\omm+h\om{0i}+\frac{1}{2}m\omT^2 x_i^2~,\label{eq:sqrtNum}
\end{equation}
where  $\om{00}=0$. From this we obtain the expression for the localized state position
$x_i =  \sqrt{2h(\nurm{mod}-\nu_{0i})/m\omT^2}$.
Fitting the data of Fig.~\ref{fig:transp_outlier_pos_vs_freq} to this functional dependence allows us to attribute the upper and lower branch to the first and second bands respectively.
From the fit we obtain $\nu_{01}^\mathrm{fit}=18.5(2)$~kHz and $\nu_{02}^\mathrm{fit}=33.06(17)$~kHz in good agreement with first ($18.1$~to~$19.3$~kHz) and second ($30.6$~to~$37.2$~kHz) Bloch bands at $s=16$. This agreement confirms that the process is well understood and demonstrates its applicability for the production of localized clouds in the lattice at a long distance from the central component.



\begin{figure}[t]
\centering
\includegraphics[width=8cm]{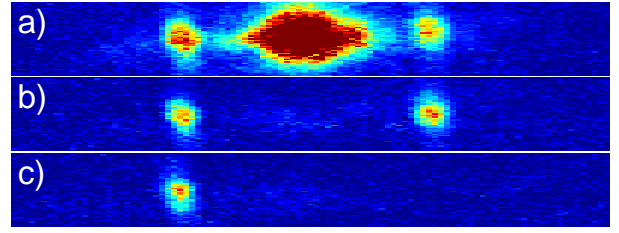}
\caption{(Color online) Individual addressing of localized states. a) Population of localized states after a modulation for $10$~ms at $30$~kHz. RF-removal of b) the lowest and c) the main component.}
\label{fig:transp_rf_cut_incr}
\end{figure}

The efficient population of localized lattice states offers avenues for e.g. precision interferometric experiments. This requires the ability to manipulate the localized states independently. To demonstrate this ability, we have
implemented selective removal of components
as illustrated in Fig.~\ref{fig:transp_rf_cut_incr}. Due to the gravitational potential, the condensate is shifted slightly downwards compared to the magnetic field zero. The gravitational sag in the weak magnetic confinement $x_{\mathrm{sag}} = g/\omT^2 = \unit{173}{\mu m}$ 
is large enough to ensure that each cloud is located at unique magnetic fields. Selected peaks can thus be removed by applying the appropriate RF-field to couple them to the unbound magnetic states. This represents an ideal starting position for future investigations of the coherence properties of the localized states.
It also enables a new set of experiments that probe non-linear matter wave dynamics, including the reflection and transmission on a BEC~\cite{Poulsen2003}, the collision of wave packets, and the investigation of their dispersive properties.

In conclusion, the experimental realization of quasi-free wave packets of ultra-cold atoms based on excitations in a combined optical lattice and harmonic trap was demonstrated. Furthermore, a probe pulse in direct analogy with pump-probe spectroscopy allowed for the resonant de-excitation of the wave packets into localized lattice states at a controllable distance of more the 100 lattice sites from the main component. Finally, the independent manipulation of the population of the localized states was implemented.

A particularly promising experimental extension will be the use of multiple modulation frequencies. On the one hand this will allow for a full analysis of the Franck-Condon type overlap between bound and quasi-continuous states. On the other hand it will enable the controlled population of many, spatially separated lattice states.
It will also be interesting whether the transport can be extended to 2D and 3D lattices and combined with single site addressing~\cite{Weitenberg:2011} provide the basis of a new model of scalable quantum computation. In addition, our preparation of localized states in selected bands may serve as a particularly pure source for the important investigation of the rich structure of extended Bose-Hubbard Hamiltonians including e.g. the supersolid state~\cite{Scarola2005}.
The precise control of the external degrees of freedom thus makes these wave packets a versatile tool  both for fundamental investigations and applications in the field of quantum technology.

We thank the Danish National Research Foundation for support 
and
J.F.S. acknowledges  support from the Danish Council for Independent Research. 

\bibliographystyle{prsty}
\bibliography{bibfile}

\end{document}